# Gapped Dirac materials and quantum valley currents in dual-gated hBN/bilayer-graphene heterostructures

Takuya Iwasaki[1], Yoshifumi Morita[2], Kenji Watanabe[3], and Takashi Taniguchi[1]

[1]*Research Center for Materials Nanoarchitectonics, National Institute for Materials Science (NIMS), 1-1 Namiki, Tsukuba, Ibaraki 305-0044, Japan*
[2]*Faculty of Engineering, Gunma University, Kiryu, Gunma 376-8515, Japan*
[3]*Research Center for Electronic and Optical Materials, NIMS, 1-1 Namiki, Tsukuba, Ibaraki 305-0044, Japan*

**Abstract**

In gapped Dirac materials, the topological current associated with each valley can flow in opposite directions creating long-range charge-neutral valley currents. We report valley currents in hBN/bilayer-graphene heterostructures with an energy gap, which is tunable by a perpendicular electric (displacement) field in a dual-gated structure. We observed significant nonlocal resistance, consistent with the scaling theory of the valley Hall effect. In the low-temperature limit, the nonlocal resistance approaches a saturated value near the "quantum limit," indicating the emergence of quantum valley currents.

**Main text**

## 1. INTRODUCTION

A valley (degenerate local minimum/maximum in the conduction/valence band, respectively) is a quantum-mechanical degree of freedom built into electrons for several solid-state systems, which is referred to as *K* and *K'* in the case of graphene [1]. In the context of graphene, the valley degree of freedom is sometimes referred to as "flavor" combined with the spin degree of freedom.

Graphene, a monolayer of carbon atoms, is the parent of novel low-dimensional quantum metamaterial [2]. Single-layer graphene (SLG) has relativistic energy bands with a linear dispersion in the low-energy limit, whereas its bilayer counterpart, AB-stacked (Bernal) bilayer graphene (BLG), has degenerate energy bands with a parabolic energy touching. Both SLG and BLG (and their heterostructures with hexagonal boron nitride (hBN) [3,4]) belong to a "Dirac-material family", although the detailed character of each material varies even within the family. Gapped Dirac materials can exhibit topological current transverse to the applied electric field even without a magnetic field or broken time-reversal symmetry. In the case of graphene, the topological current associated with each valley of an energy band can flow in opposite directions. This phenomenon is known as the valley Hall effect (VHE), which generates long-range charge-neutral valley currents. The electrical VHE was first demonstrated for hBN/SLG superlattices [5,6]. The VHE has also been observed in BLG under a perpendicular electric field [7,8]. In contrast to SLG, BLG provides an ideal platform for detailed study of valley currents because the energy gap and band structure can be systematically tuned by a perpendicular electric field [9]. The topological valley current was also observed in hBN/BLG superlattices [10]. A more recent study reported the observation of valley currents with a systematic control of the crystallographic stacking angle between hBN and BLG [11].

Here, we focus on dual-gated hBN/BLG heterostructures in which the energy gap is further tuned by applying a perpendicular displacement field. We demonstrate the detection and manipulation of the valley current with a tunable displacement field. The scaling analysis indicates that the VHE is established in the temperature range below the band gap. In the low-temperature limit, the nonlocal resistance approaches the "quantum limit", implying the emergence of quantum valley currents [6].

## 2. METHODS

To fabricate the devices, BLG and hBN flakes were first prepared on a $SiO_2$/heavily doped Si substrate by mechanical exfoliation from bulk crystals. The hBN/BLG/hBN heterostructure was assembled by a dry transfer method [12]. The top gate (Ti/Au) was fabricated via electron beam (EB) lithography and EB deposition. The heterostructure was then patterned into a Hall bar geometry using EB lithography and reactive ion etching ($CHF_3/O_2$ plasma). The edge contact electrodes (Cr/Au) were likewise fabricated via EB lithography and EB deposition [13].

To investigate the transport properties, we adopted a four-terminal configuration with AC lock-in techniques. The device was measured in a $^4$He cryostat with a variable temperature insert to control the temperature ($T$). A superconducting magnet was used to apply a magnetic field ($B$) perpendicularly to the BLG plane.

The schematic cross-section of our device is shown in Fig. 1(a). Our devices comprise the hBN/BLG/hBN heterostructure with the top and the bottom gates (so-called "dual-gated structure"), where the hBN layers play the role of a high-quality dielectric on both sides. The thicknesses of the top and bottom hBN layers are 35 nm and 34 nm, respectively, which is confirmed by an atomic force microscope. Highly doped Si was used as a back gate. By applying top-gate ($V_{tg}$) and back-gate voltage ($V_{bg}$) to the BLG, the perpendicular displacement field ($D$) and carrier density ($n$) are independently controlled, where $D = [C_{bg}(V_{bg} - V_{bg,0}) - C_{tg}(V_{tg} - V_{tg,0})]/2\varepsilon_0$, and $n = [C_{bg}(V_{bg} - V_{bg,0}) + C_{tg}(V_{tg} - V_{tg,0})]/e$; $C_{bg(tg)}$ is the back/top gate capacitance per unit area, $V_{bg,0(tg,0)}$ is the offset from the charge neutrality point (CNP), $\varepsilon_0$ is the vacuum permittivity, and $e$ is the elementary charge.

We fabricated two Hall bar devices D1 and D2, both of which yield fundamentally consistent results. In the main text, we focus on the characteristics of D1 (see Supplemental Materials S2 for D2 [14]). The optical image of D1 is displayed in the inset of Fig. 1(c). The channel length and width of D1 are $L = 2.5$ μm and $W = 1.6$ μm, respectively.

## 3. RESULTS AND DISCUSSION

Figure 1(b) shows the intensity map of the longitudinal resistivity $\rho_{xx}$ as a function of $V_{bg}$ and $B$ at $T$ = 1.7 K, where $\rho_{xx} = V_{65}/I_{14} \times (W/L)$ (see also the inset in Fig. 1(d)), $V_{kl}$ is the voltage drop between terminals k–l, and $I_{ij}$ is the current injected between terminals i–j. The peak with a high $\rho_{xx}$ at $V_{bg} \sim -0.78$ V corresponds to the CNP. We observe a typical Landau quantization spectrum of BLG fanning out from the CNP [15]. The carrier mobility estimated from Hall measurements at $T$ = 1.7 K is ~23 $m^2V^{-1}s^{-1}$ for electrons and ~19 $m^2V^{-1}s^{-1}$ for holes. The residual carrier density is ~3.6 × $10^{10}$ $cm^{-2}$, which is estimated from the full-width-at-half-maximum of the CNP peak [16]. These data indicate that our device is in an ultra-clean regime (see Supplemental Materials S1 for the high-quality properties of D1 in more detail, including the mean free path and the quantum Hall effect [14]).

The valley currents manifest themselves through nonlocal transport properties in the Hall-bar geometry (see the inset of Fig. 1(e)). In BLG with an energy gap, near the valleys of the energy bands, a finite Berry curvature emerges with an opposite sign in each valley [1]. The Berry curvature, playing the role of a (pseudo-)magnetic field in the momentum space, induces anomalous velocity with an opposite direction to electrons in each valley. Therefore, in nonlocal measurements, a transverse neutral valley current is generated by the electric current between the left-side terminals (VHE). This valley current is converted into a voltage drop between the right-side terminals (inverse VHE). As illustrated in Fig. 1(c), a finite nonlocal resistance $R_{nl} = V_{53}/I_{62}$ (see also the inset in Fig. 1(e) for the measurement configuration) is observed near the CNP of the local resistivity ($\rho_{xx}$). In general, nonlocal resistance can be attributed to the contribution from stray charge currents via the van-der-Pauw formula $R_{Ohm} = (\rho_{xx}/\pi)\exp(-\pi L/W)$. In our device, such a contribution is negligible, as shown in Fig. 1(c). The $R_{nl}$ peak is sharper than that of $\rho_{xx}$ and is consistent with previous studies on the VHE [5–8,10]. Combined with the scaling analysis below, we attribute the nonlocal response observed in our device to the emergence of the valley current.

For the local measurement configuration, the increase of $\rho_{xx}$ with an application of $D$ is observed in Fig. 1(d), which corresponds to a gap-opening at the CNP in BLG [9]. The asymmetric increase in $\rho_{xx}$ depending on the polarity (sign) of $D$ can be attributed to different dielectric environments between the top and back gates in our device. For the nonlocal measurement configuration shown in Fig. 1(e), the $R_{nl}$ peak appears for $|D|$ > 6 mV/nm. The maximum $R_{nl}$ increases with larger $|D|$ and reaches an order of kΩ at $D \sim 27$ mV/nm (see also Fig. 2(b) to confirm the numeric relationship of $D$ vs. $R_{nl}$). As shown in Ref. [7,8], without an alignment between hBN and BLG, the nonlocal resistance is strongly suppressed to a near-zero value in a small displacement-field regime. To be more precise, the nonlocal resistance cannot be distinguished from the small Ohmic contribution. In our device D1, the nonlocal resistance is enhanced even under such a small displacement field. In addition, in another device D2, the data show a finite nonlocal resistance even under zero displacement field (see Supplemental Materials S2). The device D2 is close to D1 in the alignment angle since they are from the same stack. The increase of the $R_{nl}$ with a different polarity in $D$ is also asymmetric here, but exhibits an opposite behavior to that of $\rho_{xx}$. Detailed study of these trends is left as a future task. As commented above, different dielectric environments between the top and back gates may play some role here. The sign of $D$ should correspond to the difference in the shift of low-energy electronic states toward (or away from) the top and bottom hBN's with different settings (width, alignment etc.). Compared with the nonlocal resistance of conventional BLG for the same $D$ [7,8], the threshold of $D$ for a finite $R_{nl}$ is smaller, and the maximum value of $R_{nl}$ is much higher in our device, approaching the "quantum limit" as discussed below. The $R_{nl}$ reaches a kΩ order at maximum, indicating that the nonlocal resistance is approaching the quantum limit [6]. At the quantum limit, the nonlocal resistance illustrates an order of $\sim h/4e^2$ ($h$ is the Planck constant) apart from a prefactor of order 1, which also implies a large valley Hall angle. In this study, as discussed in Ref. [6], we ascribe this to "quantum" valley currents.

Let us now discuss the $T$-dependence. In Figs. 2(a) and (b), the $T$-dependence of maximum $\rho_{xx}$ ($\rho_{xx,max}$) and $R_{nl}$ ($R_{nl,max}$) demonstrates a thermally activated behavior in the high-$T$ regime. The $T$-dependence of $\rho_{xx,max}$ and $R_{nl,max}$ becomes weak in the low-$T$ regime, where hopping conduction dominates. Moreover, the temperature at which the conduction mechanism switches between thermal activation and hopping conduction becomes higher for larger $|D|$. This $T$-dependence is consistent with previous studies on BLG [7,8]. Using the Arrhenius fit, i.e., $1/\rho_{xx,max}$ or $1/R_{nl,max} \sim \exp(-E_g/2k_BT)$ (where $E_g$ is the energy gap, $k_B$ is the Boltzmann constant) for the high-$T$ regime (insets in Figs. 2(a) and (b)), the energy gaps for local ($E_{g,local}$) and nonlocal configuration ($E_{g,nl}$) are extracted and summarized in Fig. 2(c). For both local and nonlocal configurations, $E_g$ increases with the application of finite $|D|$. At $D = 0$, $E_g$ is ~11 K and comparable with previous reports on different hBN/BLG superlattices with a small misalignment [10]. This is consistent with the observation of the "Umklapp effects" in resistivity (discussed in Supplemental Materials S3 [14]). The finite energy gap and the Umklapp effects are absent with a large misalignment between hBN and BLG [17] (see Supplemental Materials S4 for the alignment between hBN and BLG in our device [14]). At $D = 0$, $E_g$ is well-defined only for the local configuration, since $R_{nl,max} \sim 0$ and the energy gap is difficult to define here (Fig. 1(e)). When the local conductivity ($\sigma_{xx} = \rho_{xx}^{-1}$) exceeds the valley Hall conductivity ($\sigma_{xy}^v$), i.e., $\sigma_{xx} \gg \sigma_{xy}^v$, the nonlocal resistance is described by the following formula [18]:

$$R_{nl} = \frac{W}{2l_v}(\sigma_{xy}^v)^2 \rho_{xx}^3 \exp\left(-\frac{L}{l_v}\right), \tag{1}$$

where $l_v$ is the intervalley scattering length. Thus, $E_{g,nl} \sim 3E_{g,local}$ is expected. In our data, $E_{g,nl}$ is approximately three times larger than $E_{g,local}$ for $D \geq 18.3$ mV/nm (e.g., $E_{g,nl} \sim (2.88\pm0.05)E_{g,local}$ at $D = 18.3$ mV/nm, $E_{g,nl} \sim (2.72\pm0.07)E_{g,local}$ at $D = 27.3$ mV/nm, the error is estimated by the Arrhenius fitting), which demonstrates a consistent trend.

Next, we discuss the scaling relation between $R_{\text{nl,max}}$ and $\rho_{xx,\text{max}}$. As expected from Eq. (1), the cubic scaling relation between $R_{\text{nl,max}}$ and $\rho_{xx,\text{max}}$ is found in the range $T = 12$–$25$ K at $D = 27.3$ mV/nm (Fig. 3(a)). In the higher-$T$ regime, the valley Hall conductivity should exhibit a deviation from the quantum value, and it is reasonable that the cubic scaling does not hold. In the low-$T$ limit, the nonlocal resistance can approach a saturated value in the quantum limit. Moreover, our data for $D \geq 18.3$ mV/nm display the cubic scaling relation in the intermediate-$T$ regime as illustrated in Fig. 3(b). This $D$ range is the same as one for which $E_{g,\text{nl}} \sim 3E_{g,\text{local}}$ holds (Fig. 2(c)), as discussed above. In the $T$ range below the $E_{g,\text{local}}$, it is reasonable to assume $\sigma_{xy}^{v} = 4e^2/h$ at the CNP [11]. Plugging this $\sigma_{xy}^{v}$ into Eq. (1), the nonlocal resistance can be described by $R_{\text{nl}} = \rho_{xx}^{p}(W/2l_v)(4e^2/h)^2\exp(-L/l_v)$, where $p$ is ideally 3 and actually, we treat it as a fitting parameter. By fitting this formula to our data in the range where the cubic scaling relation holds, i.e., $p \sim 3$ (Fig. 3(c)), we obtain $l_v = 2.5$ μm, which is near the scale of our device and compatible with a previous report (e.g., [5]).

Comments are in order on some ambiguity in the weak-$D$ regime; we note that, when the gap is small, disorder effects can obscure a clear scaling relation even in ultra-clean devices and, moreover, a precursor toward possible phase transition/instability can play some role. In our recent work [19], we revealed that in the phase diagram of hBN/BLG superlattices the weak-$D$ regime is a "bifurcation" point of phase boundaries, where some theories ascribe such instabilities to competing orders [20–23].

Finally, let us comment more on the "quantum limit" terminology [6]. For $\sigma_{xx} \ll \sigma_{xy}^{v}$ (the valley Hall angle $\sim \pi/2$), $R_{\text{nl}} = (W/2l_v)h/4e^2$ was proposed [24]. In real settings, empirically in ultra-clean devices, $W$ and $l_v$ are of the same order, which is also reconfirmed in our setting. As discussed above, the cubic scaling was verified in the regime where the $R_{\text{nl}}$ is small (small/moderate valley Hall angle). On the other hand, when the valley Hall angle is large, the nonlocal resistance can exhibit an order of $\sim h/4e^2$, i.e., approaching the quantum limit near $R_{\text{nl}} \sim 2.1$ kΩ for this device. In Ref. [6], a possible scenario of the edge-mode conduction was proposed deeply inside the quantum limit. Since our outputs are still below the genuine quantum limit, we presume that bulk conduction still dominates the transport properties in this regime and the scaling relations hold. On the other hand, when the scaling relation breaks, a possible scenario is the edge-conduction picture [6].

## 4. SUMMARY

We investigated the local and nonlocal transport properties in dual-gated hBN/BLG heterostructures, which belong to gapped Dirac materials. The observation of giant nonlocal resistance is consistent with the scaling theory of the valley Hall effect. In the low-temperature limit, on the other hand, we observed nonlocal resistance near the quantum limit. To elucidate the conduction mechanism ("edge vs. bulk" etc.) in this low-temperature limit, further experiments with multi-terminal devices and theoretical modeling are crucial. Our work should lay a sound basis for next-generation devices based on quantum metamaterials with nano-structures like quantum dot, point contact and their hybrids.

**ACKNOWLEDGMENTS**

The authors thank H. Osato, E. Watanabe, and D. Tsuya from the NIMS Nanofabrication Facility for discussing the device fabrication. This work was partially supported by JPSJ KAKENHI Grant No. 21H01400, and "Advanced Research Infrastructure for Materials and Nanotechnology in Japan (ARIM)" of the Ministry of Education, Culture, Sports, Science and Technology (MEXT). Proposal Number JPMXP1223NM5186.

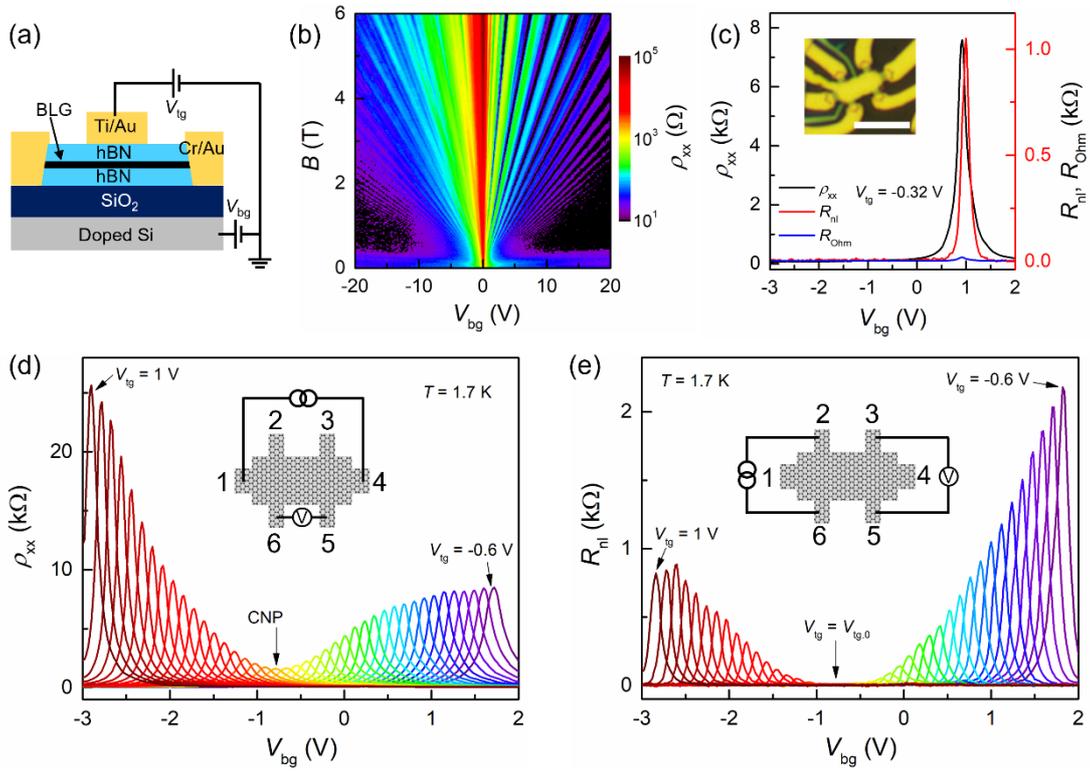

FIG. 1. (a) Schematic cross-section of the device. (b) Landau fan diagram: the intensity map of $\rho_{xx}$ as a function of $V_{bg}$ and $B$ at $T = 1.7$ K and $V_{tg} = 0$ V. (c) $\rho_{xx}$ (black), $R_{nl}$ (red), and $R_{Ohm}$ (blue) as a function of $V_{bg}$ at $T = 1.7$ K, $B = 0$ T, and $V_{tg} = -0.32$ V. The inset shows the optical image of the device. The scale bar corresponds to 5 μm. (d) $\rho_{xx}$ as a function of $V_{bg}$ for $V_{tg}$ from $-0.6$ V to 1 V. The inset shows the schematic of the local measurement configuration. (e) Same plot as (d) for the nonlocal configuration.

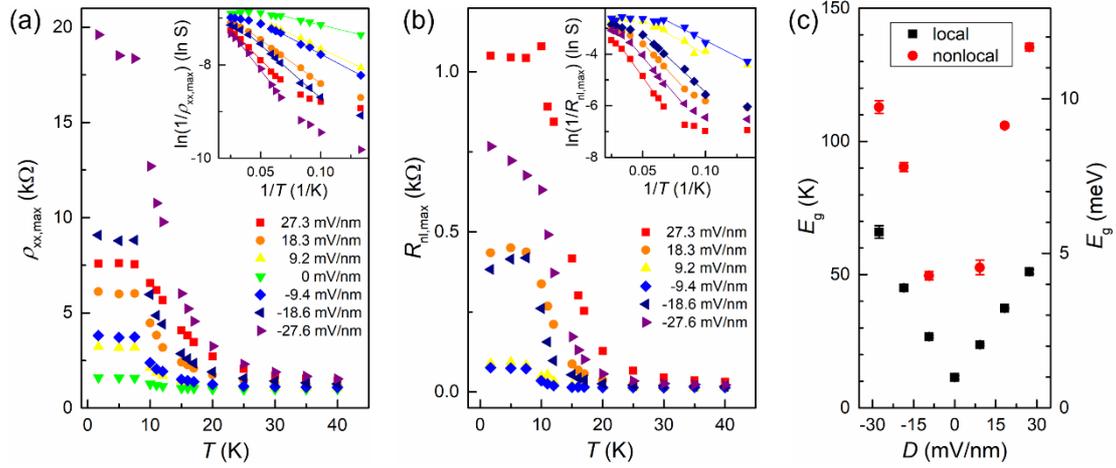

FIG. 2. (a,b) Temperature dependence of (a) $\rho_{xx}$ and (b) $R_{nl}$ for various $D$. The insets show the Arrhenius plots for (a) $\rho_{xx,max}$ and (b) $R_{nl,max}$, respectively. The solid lines show the fitting to $\sim\exp(-E_g/2k_BT)$. (c) Energy gap extracted from the Arrhenius fit for $\rho_{xx,max}$ (local, black) and $R_{nl,max}$ (nonlocal, red) as a function of $D$. The error bars correspond to the ambiguity in the fitting procedure.

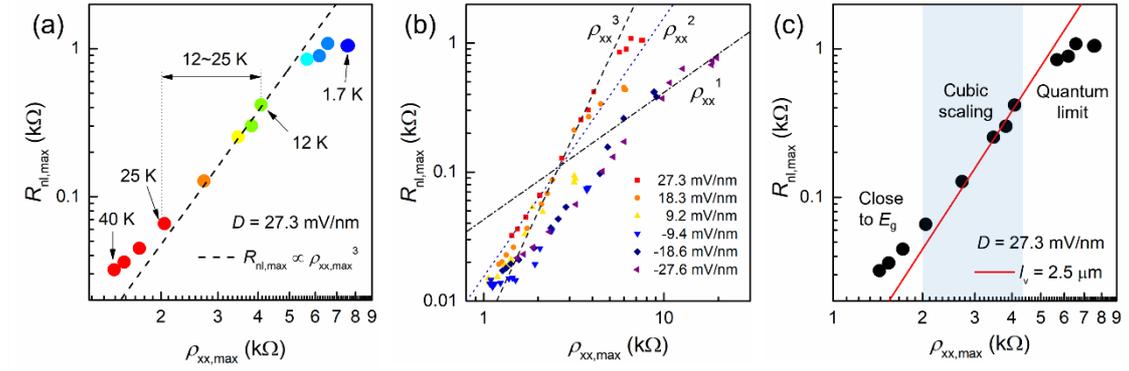

FIG. 3. Scaling analysis in the log-log plot between $\rho_{xx,max}$ and $R_{nl,max}$. The data are obtained from each $T$ ranging 1.7-40 K. (a) At $D = 27.3$ mV/nm. The dashed line represents a fitting to the cubic relation ($\sim\rho_{xx,max}^3$). (b) The same as (a) with various $D$. The dashed, dotted, dot-dashed lines correspond to the relation $\sim\rho_{xx,max}^3$, $\sim\rho_{xx,max}^2$, $\sim\rho_{xx,max}^1$, respectively. (c) At $D = 27.3$ mV/nm. The solid line illustrates the fitting result with $p = 3.09$, $L = 2.5$ μm, $W = 1.6$ μm, and $l_v = 2.5$ μm.

# Supplemental Materials for
## "Gapped Dirac materials and quantum valley currents in dual-gated hBN/bilayer-graphene heterostructures"

Takuya Iwasaki[1], Yoshifumi Morita[2], Kenji Watanabe[3], and Takashi Taniguchi[1]

[1]*Research Center for Materials Nanoarchitectonics, National Institute for Materials Science (NIMS), 1-1 Namiki, Tsukuba, Ibaraki 305-0044, Japan*
[2]*Faculty of Engineering, Gunma University, Kiryu, Gunma 376-8515, Japan*
[3]*Research Center for Electronic and Optical Materials, NIMS, 1-1 Namiki, Tsukuba, Ibaraki 305-0044, Japan*

**S1. High-quality characteristics of the device D1**
**S2. Characterization of another device D2**
**S3. Temperature dependence of the resistivity away from the charge neutrality point**
**S4. Alignment between hBN and BLG**

## S1. High-quality characteristics of the device D1

In the device D1 of the main text, the quantum Hall effect is observed (Fig. S1 (a)); quantum Hall plateaus are observed with the quantized value of the Hall resistivity $\rho_{xy} = V_{53}/I_{14}$, when the longitudinal resistivity $\rho_{xx}$ exhibits minima. In particular, at the filling $\nu = 4N$ ($N$ is an integer), the $\rho_{xx}$ is strongly suppressed and the $\rho_{xy}$ plateaus become pronounced. Furthermore, some degeneracies are lifted, and additional plateaus also occur, for example, $\nu = 2$. The residual carrier density $n_{res}$ is a probe of the inhomogeneity in the device [1]. In our device D1, $n_{res}$ (~3.6 × 10$^{10}$ cm$^{-2}$) is estimated by the full-width-at-half-maximum of the peak in the $\rho_{xx}$-carrier density ($n$) plot (Fig. S1(b)). The mean free path $(h/2e^2)(1/\rho_{xx}\sqrt{\pi n})$ approaches the scale of the device geometry (Fig. S1(c)). The carrier mobilities estimated from the Hall measurement surpass those of the conventional BLG devices [2,3] (Fig. S1(d)). These features guarantee the ultra-high quality of our device.

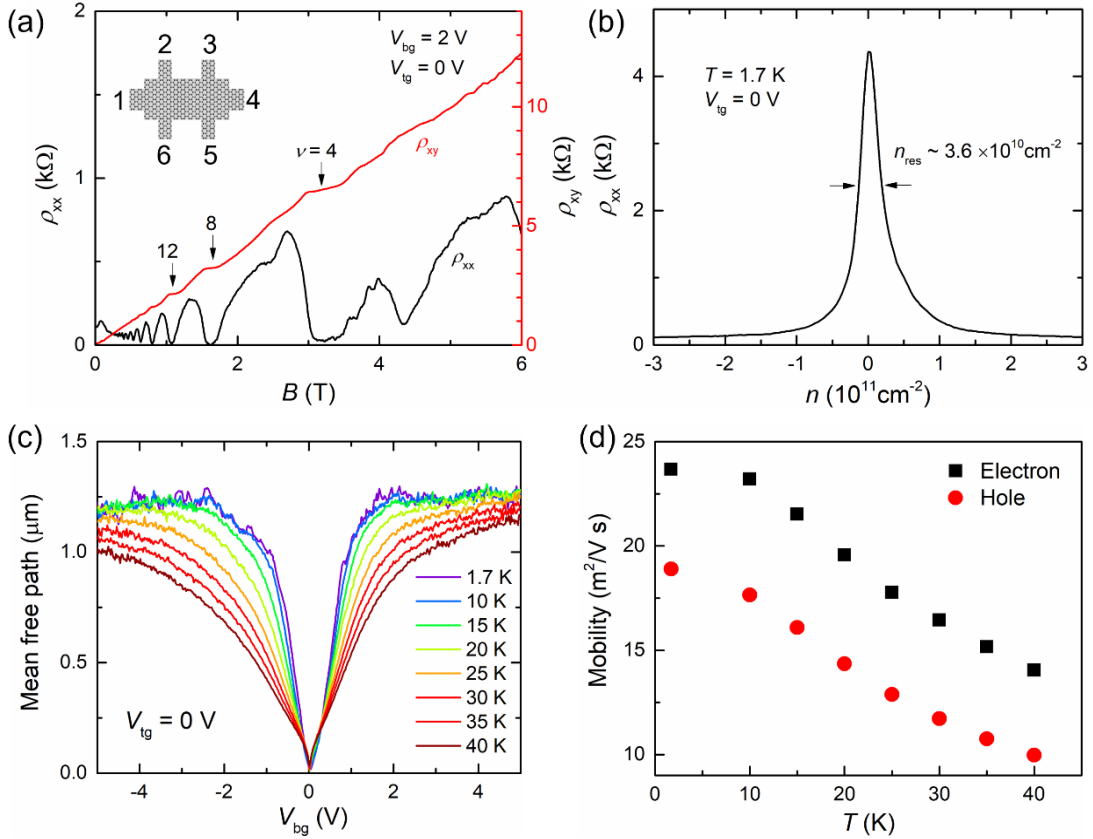

Fig. S1. High-quality characteristics of the device D1 discussed in the main text. (a) $\rho_{xx}$ (black) and $\rho_{xy}$ (red) as a function of a perpendicular magnetic field $B$ at the temperature $T = 1.7$ K, $V_{bg} = 2$ V, and $V_{tg} = 0$ V. The inset shows the schematic of the device geometry. (b) $\rho_{xx}$ as a function of $n$ at $T = 1.7$ K and $V_{tg} = 0$ V. (c) Mean free path as a function of $V_{bg}$ for various $T$. (d) $T$-dependence of the carrier mobility estimated from the Hall measurement.

# S2. Characterization of another device D2

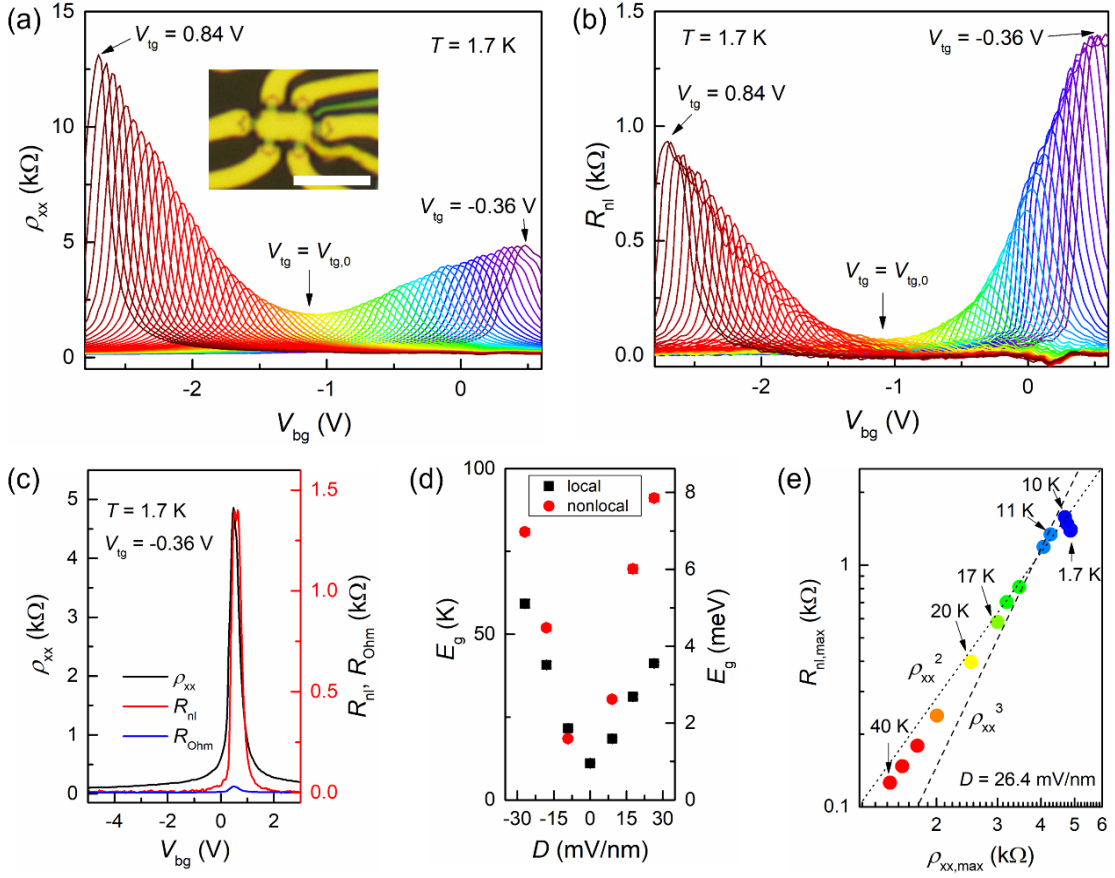

Fig. S2. Characterization of device D2. (a) $\rho_{xx}$ as a function of $V_{bg}$ for $V_{tg}$ from –0.36 V to 0.84 V. The inset shows the optical image of D2. (b) Same plot as (a) for the nonlocal resistance $R_{nl}$. (c) $\rho_{xx}$ (black), $R_{nl}$ (red), and $R_{Ohm}$ (blue) as a function of $V_{bg}$ at $T = 1.7$ K, $B = 0$ T, and $V_{tg} = –0.36$ V. (d) $E_g$ as a function of $D$. $E_g$ is extracted from the Arrhenius fit for the local (black) and nonlocal configuration (red). (e) Log-log plot between $\rho_{xx,max}$ and $R_{nl,max}$. The data are obtained from 1.7 K to 40 K at $D = 26.4$ mV/nm. The dotted and dashed lines show a fitting to the quadratic and cubic scaling relations, respectively.

We also fabricated another device D2. Although D2 is close to D1 in the alignment angle between hBN and BLG, the quantitative character is different due to the angle-disorder etc. even in the same stack. The optical image of D2 is displayed in the inset of Fig. S2(a). The channel length and width of D2 are $L = 2.0$ μm and $W = 1.6$ μm, respectively. For the local measurement configuration, the increase of $\rho_{xx}$ with the application of finite displacement field $D$ is observed (Fig. S2(a)). For the nonlocal measurement configuration shown in Fig. S2(b), the $R_{nl}$ peak appears even at $D = 0$ ($V_{tg} = V_{tg,0} = 0.24$ V, $V_{bg} = V_{bg,0} = –1.1$ V). The maximum $R_{nl}$ is in the order of kΩ at $D \sim 21$ mV/nm under a displacement field. As shown in Fig. S2(c), the peak of $R_{nl}$ is sharper than that of $\rho_{xx}$ and the Ohmic contribution to $R_{nl}$ is minor. The energy gap is deduced from the Arrhenius fitting to the $T$-dependence of $\rho_{xx,max}$ and

$R_{nl,max}$ (Fig. S2(d)). The energy gaps do not satisfy the relation $E_{g,nl} \sim 3E_{g,local}$ in this $D$ range as in the small-$D$ regime of device D1. Furthermore, as shown in Fig. S2(e), at $D = 26.4$ mV/nm, the scaling relation between $\rho_{xx,max}$ and $R_{nl,max}$ does not show a clear cubic behavior and the proper scaling regime is narrow, if any. At the low-$T$ limit, $R_{nl}$ tends to be saturated near an order of $\sim h/4e^2$, i.e., approaching the quantum limit.

**S3. Temperature dependence of the resistivity away from the charge neutrality point**

In the main text, our focus is on the nonlocal (and local) transport properties at CNP. In these supplementary data, to be complementary with the main text, local transport properties away from the CNP are provided. The nonlocal transport away from the CNP is strongly suppressed as shown in Fig. 1(e) in the main text and Fig. S2(b). Figure S3(a) shows $\rho_{xx}$ as a function of $V_{bg}$ with $V_{tg} = 0$ in device D2. Near the CNP ($D \sim 0$), an insulating behavior is observed as demonstrated in Fig. S3(f). Although it is reasonable to assign this to an energy gap induced by the moiré effect, interaction-driven phase transition in BLG can happen with a spontaneous generation of the energy gap, where "flavor"-symmetry etc. are broken [4–8]. But no such a spontaneous gap generation is established in hBN/BLG heterostructures and we consider that the gap is induced by the moiré effect. When the carrier is doped from the CNP ($n = 0$), the $\rho_{xx}$–$T$ characteristic changes to a metallic behavior with a strong $T$-dependence. This is in contrast to the previous study on the BLG/hBN heterostructure, where the $T$-dependence is very weak even at a highly-doped regime [9]. To extract the power-law dependence in the hBN/BLG superlattice, we performed a fitting of the data to $\Delta\rho_{xx} \sim \alpha T^\beta$ ($\Delta\rho_{xx} = \rho_{xx}(T) - \rho_{xx}(T = 1.7$ K), $\alpha$ and $\beta$ are fitting parameters) for the high-$T$ regime ($T = 70 \sim 300$ K) as shown in Figs. S4(a) and S4(b). The exponent $\beta$ varies between $\sim 1$ and $\sim 2$ (see the inset of Fig. S4(b)). In the pristine graphene, the $T$-dependence of the resistance is essentially caused by phonons, which leads to the $T$-linear dependence of $\rho_{xx}$ in such a high-$T$ regime [10,11]. In contrast, electron-electron interaction (to be more precise, "Umklapp effects" due to superlattice structure caused by the moiré effect) can dominate the scattering process in hBN/graphene superlattices, leading to the $T^2$-dependence of $\rho_{xx}$ [12–14]. In doubly aligned hBN/BLG/hBN superlattices, a similar crossover was reported between $\beta$ $\sim 1$ and $\sim 2$ [15]. More detailed assessment of the scattering mechanism in our hBN/BLG heterostructures is left as a future task, which should take into account critical fluctuations of possible competing orders, including scenarios of ref. [4–8] and beyond.

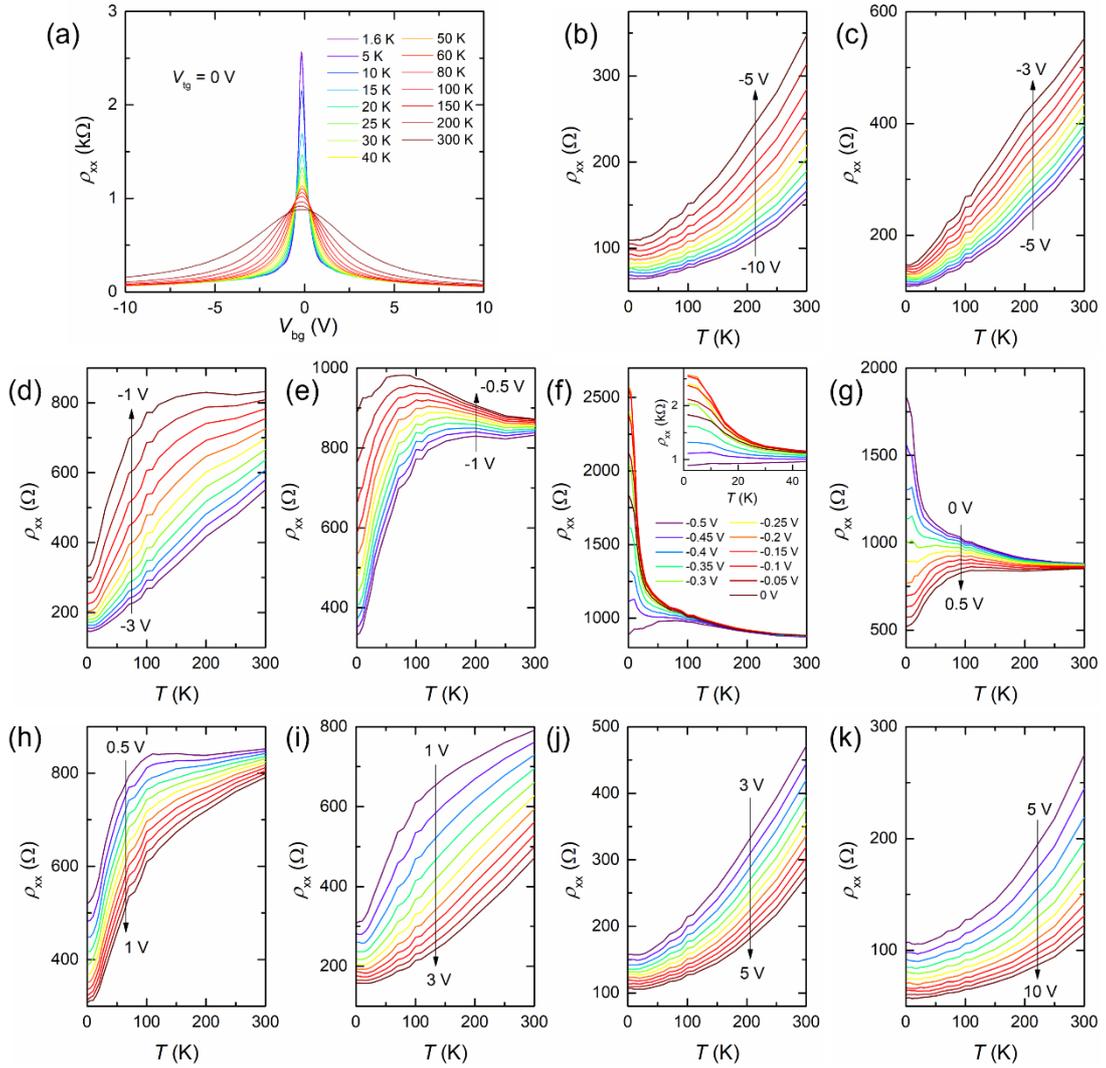

Fig. S3. Transport properties away from the CNP in device D2. (a) $\rho_{xx}$ as a function of $V_{bg}$ for various $T$ at $V_{tg} = 0$ V. (b–k) $T$-dependence of $\rho_{xx}$ for various $V_{bg}$. (b) $V_{bg} = -10 \sim -5$ V. (c) $V_{bg} = -5 \sim -3$ V. (d) $V_{bg} = -3 \sim -1$ V. (e) $V_{bg} = -1 \sim -0.5$ V. (f) $V_{bg} = -0.5 \sim 0$ V in the vicinity of the CNP. The inset shows the zoom-up for the low-$T$ region. (g) $V_{bg} = 0 \sim 0.5$ V. (h) $V_{bg} = 0.5 \sim 1$ V. (i) $V_{bg} = 1 \sim 3$ V. (j) $V_{bg} = 3 \sim 5$ V. (k) $V_{bg} = 5 \sim 10$ V.

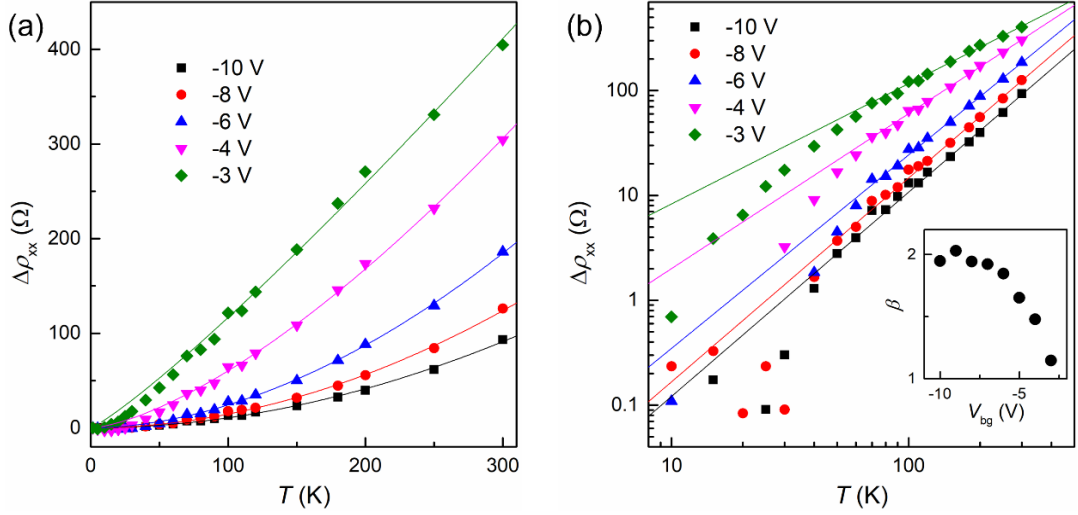

Fig. S4. Power-law $T$-dependence of $\rho_{xx}$ away from the CNP. (a) Temperature-dependent part of the longitudinal resistivity as a function of $T$ for various $V_{bg}$ at $V_{tg} = 0$ V. (b) Log-log plot of (a). The solid lines in (a) and (b) represent the fitting with $\Delta\rho_{xx} \sim \alpha T^{\beta}$. The inset in (b) shows the exponent $\beta$ of the power-law behavior of $\rho_{xx}$ as a function of $V_{bg}$.

## S4. Alignment between hBN and BLG

More on the detail of our heterostructure, alignment between hBN and BLG, is shown in Fig. S5. After the careful alignment between hBN and BLG, we applied the annealing procedure, which led to further alignments [16].

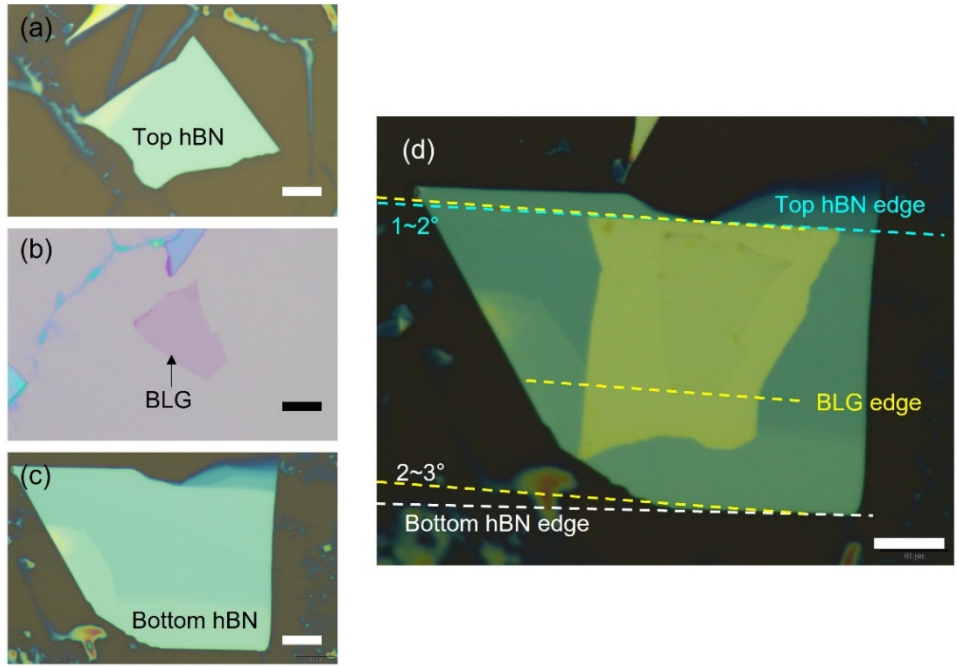

Fig. S5. Alignment between hBN and BLG in our heterostructures. (a-d) Optical images of (a) the top hBN, (b) BLG, (c) bottom hBN flakes before stacking and (d) the hBN/BLG/hBN heterostructure. The scale bar corresponds to 10 μm. The dotted lines in (d) exhibit the edge of each flake.